# Hydrocarbons under pressure: phase diagrams and surprising new compounds in the C-H system

Anastasia S. Naumova,*[†‡§] Sergey V. Lepeshkin,[†§] and Artem R. Oganov[†‡⊥]

[†] Skolkovo Institute of Science and Technology, Skolkovo Innovation Center, 3 Nobel St., Moscow, 143026, Russian Federation
[‡] Moscow Institute of Physics and Technology, 9 Institutskiy Lane, Dolgoprudny city, Moscow Region, 141700, Russian Federation
[§] Lebedev Physical Institute, Russian Academy of Sciences, 119991 Leninskii prosp. 53, Moscow, Russia
[⊥] International Center for Materials Discovery, Northwestern Polytechnical University, Xi'an, 710072, China

**ABSTRACT:** Understanding the high-pressure behavior of C-H system is of great importance due to its key role in organic, bio-, petroleum and planetary chemistry. We have performed a systematic investigation of the pressure-composition phase diagram of the C-H system at pressures up to 400 GPa using evolutionary structure prediction coupled with *ab initio* calculations and discovered that only saturated hydrocarbons are thermodynamically stable. Several stable methane-hydrogen co-crystals are predicted: $2CH_4 * H_2$, earlier obtained experimentally, is predicted to have *I*4/*m* space group and 2-90 GPa stability range at 0 K, and two new thermodynamically stable compounds $2CH_4 * 7H_2$ (*P*-3*m*1 space group) and $CH_4 * 9H_2$ (*Cm* space group), as potential energy storage materials. *P*$2_1$/*c* phase of methane is predicted to be stable at pressures < 8 GPa; bulk graphane (CH) was shown to be thermodynamically stable at 7-18 and 18-50 GPa and 0 K in the *P*-3*m* and *Cmca* phases, respectively; polyethylene is shown to have a narrow field of stability. We report the *p-T-x* phase diagram of the C-H system and *p-T* phase diagram of $CH_4$.

## INTRODUCTION

Hydrocarbons are of great interest due to their importance in many fields, such as organic chemistry, planetary science and other.[1,2] For instance, they play a key role in giant planets such as Neptune and Uranus. Today's models of these planets postulate a three-layered structure, consisting of an inner rocky core, middle ice layer of compounds of C, N, H, O and an outer H-He atmosphere.[3] Abundance of methane in the ice layer is pretty high (about 10-15 %mass, depending on the model used) at pressures up to several hundred GPa.[4] In previous works it was shown,[5-7] that ethane, butane and polymeric hydrocarbons are also likely to exist under conditions corresponding to the middle layer. However, in one theoretical paper stability of butane was questioned.[8] Later we will come back to this issue and show that chemistry of this middle layer is surprisingly diverse. Interestingly, Neptune has unexplained internal heat production responsible for its high luminosity. One of the possible explanations is that at high pressures methane decomposes into hydrogen and diamond.[6,7,9-12] The latter is denser than hydrocarbons and hydrogen and, therefore, gravitationally sinks inside the planet.[13] Most likely this process involves multiple stages and several intermediate compounds are produced during methane polymerization. In agreement with this consideration is the fact that methane, ethane and butane cannot sink and may leak into the atmosphere, where $CH_4$ and $C_2H_6$ were found indeed.[14]

Another interesting fact about C-H compounds comes from organic chemistry: benzene ($C_6H_6$) was known for many years, but its thermodynamically more stable isomer graphane[15] – $(CH)_\infty$ – was discovered experimentally just in 2009,[16] and recently bulk graphane was also synthesized.[17] Can there be other stable, but as yet unknown, hydrocarbons?

We have made an attempt to figure out this question, and found that despite its great importance, the phase diagram of the C-H system is poorly known. At ambient pressure only methane lies on the convex hull (all other hydrocarbons are thermodynamically unstable and have a higher free energy then a mixture of graphite, hydrogen and methane exothermically), above 4 GPa it was shown experimentally that it forms different co-crystals with H$_2$,[18] which normally decompose when pressure is raised beyond several GPa. Theoretical predictions show one remarkable exception: 2CH$_4$ * 3H$_2$ that was found to be stable up to 215 GPa.[19] Methane itself adopts several phases obtained experimentally, but the structures of some of them remain unidentified.[20,21] At pressures above 155 GPa pure methane was predicted to decompose into a mixture of hydrogen and ethane, and at still higher pressures butane and polyethylene were predicted to form.[5] Further pressure increase leads to the formation of diamond and hydrogen.[9] On the other hand, as a result of experimental studies methane was reported to remain stable at room temperature and compression up to 200 GPa (but note that it can be metastable at such low temperature as 300 K),[22] another static compression experiments showed its stability up to 80 GPa and 2000 K.[23]

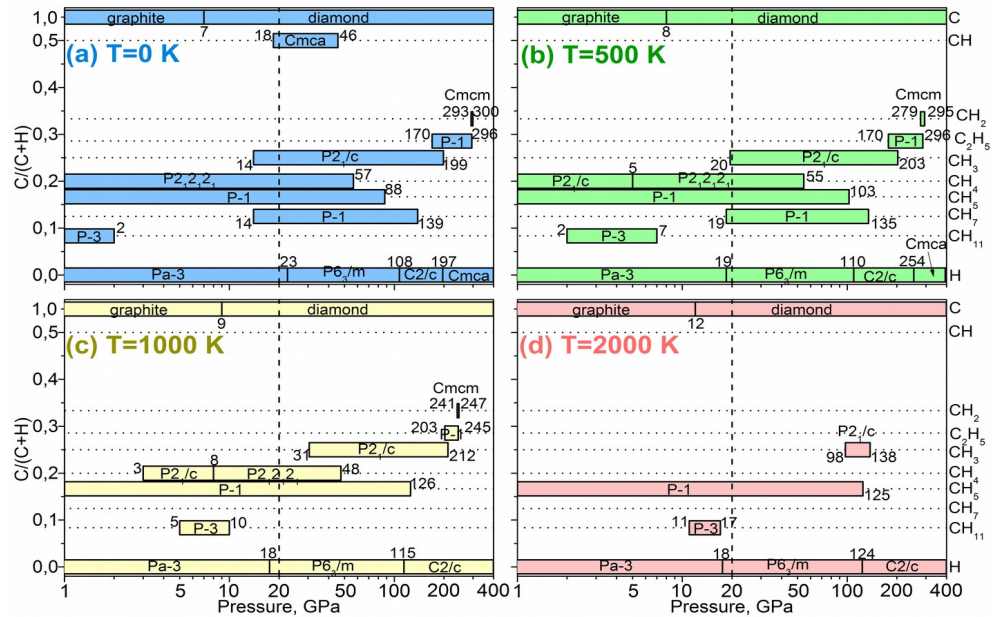

Figure 1. Pressure-composition phase diagram representation of C-H system at 0-400 GPa calculated using optB88-vdW (at 1-20 GPa) and PBE (at 20-400 GPa) functionals with zero-point energy correction included at a) 0 K, b) 500 K, c) 1000 K, d) 2000 K. CH = graphane; CH$_2$ = polyethylene; C$_2$H$_5$ = butane; CH$_3$ = ethane; CH$_4$ = methane; CH$_5$ = 4CH$_4$ * 2H$_2$; CH$_7$ = 2CH$_4$ * 3H$_2$; CH$_{11}$ = 2CH$_4$ * 7H$_2$.

Despite great importance, only one systematic study of C-H system chemical properties under pressure was done.[5] In that work authors used fixed-composition search, that didn't allow finding structures with unusual compositions, because compositions had to be manually set. We have completed studies on this area with recently developed variable-composition search, introduced in USPEX, that allows to find stable structures with any composition automatically, in combination with accurate DFT calculations,[24-26] as a result we report full phase diagram of C-H system at pressures 0-400 GPa and temperatures from 0 K up to 2000 K.

## METHODS

We use evolutionary algorithm USPEX in combination with density-functional calculations. Variable-composition method allowed us to search for all stable compounds in the C-H system. USPEX searches were performed at pressures 0, 1, 2, …10, 12, 14, …20, 40, 60, …100, 150, 200, …400 GPa and involved more than 200 000 structure relaxations. Total energy calculations and structure relaxations were performed using PBE functional in the framework of the PAW method,[27-29] with hard potentials, 850 eV plane wave kinetic energy cutoff and a uniform Γ-centered grid with $2\pi * 0.056$ Å$^{-1}$ spacing for reciprocal space sampling, optB88-vdW functional with van der Waals correction at pressures 0-20 GPa, when dispersion interactions are still important,[30] and SCAN functional with rVV10 correction and 1000 eV plane wave kinetic energy cutoff,[31] as implemented in the VASP code.[32] For all calculations with vdW corrections less dense Γ-centered grid was applied – $2\pi * 0.064$ Å$^{-1}$ – as it was done in Saleh's work.[19]

For thermodynamically stable structures phonon dispersion curves, phonon density of states and phonon contribution to the free energy were calculated using the finite displacement method as implemented in the PHONOPY code.[33] The dynamical stability of novel compounds was ascertained by the absence of imaginary frequencies in their phonon dispersion curves. Large supercells (~10x10x10 Å) and increased energy cutoff (1000 eV) were adopted in order to avoid artificial imaginary frequencies. In case of *P*-3*m*1-graphane a very large supercell (6x6x4,

15x13x16 Å, 576 atoms) was needed. The pressure-composition and $p$ – $T$ phase diagrams were obtained using the computed Gibbs free energies at given pressure and temperature in the quasi-harmonic approximation; thermal expansion is taken into account. The chosen approach is validated by a number of reference papers where phase diagrams of various materials were calculated.[34-36]

## RESULTS AND DISCUSSION

We calculated phase diagram of C-H system at 1-400 GPa with and without vdW corrections (optB88-vdW and PBE functionals, respectively) and with and without zero-point energy. Also, one more re-calculation of the whole pressure-composition phase diagram was done using SCAN functional (see supporting information for additional picture representations).[31] Results obtained with these functionals look quite similar and differ in exact stability pressure intervals, the more significant changes will be discussed below. Further in the text we will consider calculations with vdW correction up to 20 GPa, without it at pressures more than 20 GPa and always including ZP energy, unless stated otherwise. Using all predicted stable structures, we have computed phase diagrams at different temperatures: 500 K, 1000 K and 2000 K (Fig. 1).

All stable compounds discovered in previous structure prediction works were reproduced in our simulations[5-8] and a number of new stable compounds were predicted. We found that there are no stable unsaturated (including aromatic) hydrocarbons on the phase diagrams. One more interesting fact that comes from our calculations is that several methane-hydrogen co-crystals are thermodynamically stable and have wide stability range. Further we will review obtained compounds, which are newly discovered, or having some differences in comparison with previous studies (their structures are shown in fig. 2).

**Butane** $C_4H_{10}$ (*P*-1 space group): its thermodynamic and dynamical stability is debated in literature.[5,8] According to our

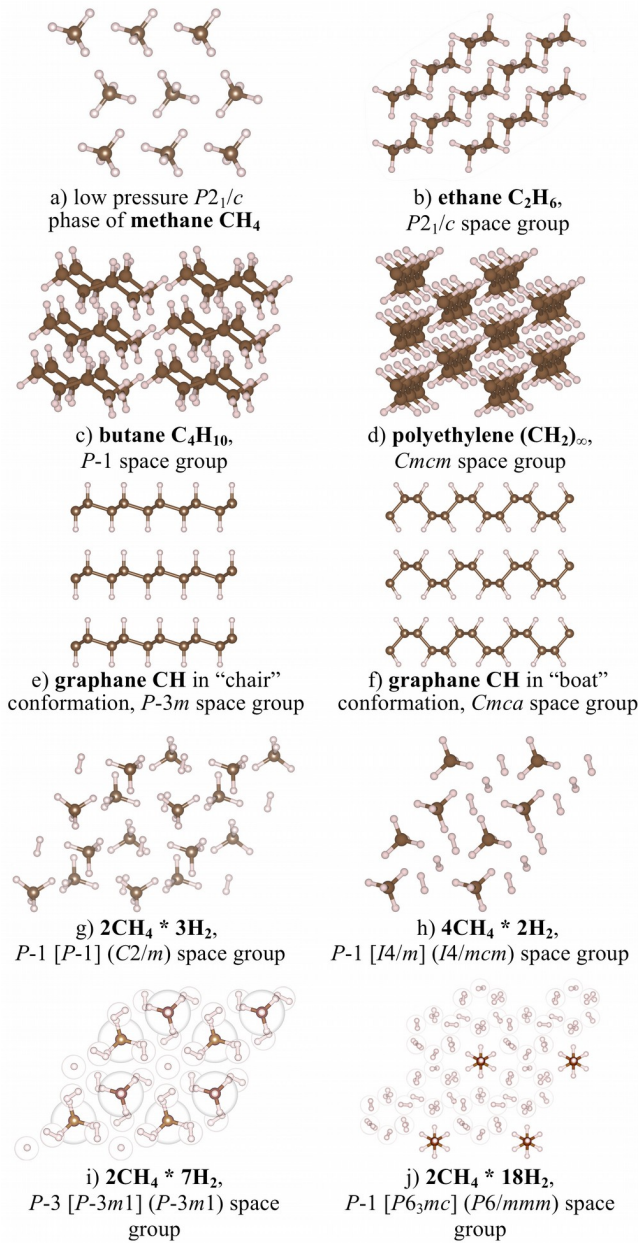

a) low pressure $P2_1/c$ phase of **methane** $CH_4$

b) **ethane** $C_2H_6$, $P2_1/c$ space group

c) **butane** $C_4H_{10}$, *P*-1 space group

d) **polyethylene** $(CH_2)_\infty$, *Cmcm* space group

e) **graphane** CH in "chair" conformation, *P*-3*m* space group

f) **graphane** CH in "boat" conformation, *Cmca* space group

g) $2CH_4 * 3H_2$, *P*-1 [*P*-1] (*C*2/*m*) space group

h) $4CH_4 * 2H_2$, *P*-1 [*I*4/*m*] (*I*4/*mcm*) space group

i) $2CH_4 * 7H_2$, *P*-3 [*P*-3*m*1] (*P*-3*m*1) space group

j) $2CH_4 * 18H_2$, *P*-1 [*P*6$_3$*mc*] (*P*6/*mmm*) space group

Figure 2. Structures of predicted compounds with their space groups. Spheres indicate spherical symmetry of the molecule inside and are shown only in those cases, where they help to determine higher symmetry. Space groups in square brackets were determined considering all molecules to be spherical (with centers in their centers of mass), in round brackets – considering only carbons as centers of mass.

results, at 0 K butane is present on the phase diagram from 170 to 296 GPa and it remains thermodynamically stable at temperatures up to 1200 K.

**Polyethylene** $(CH_2)_\infty$ has *Cmcm* space group and is thermodynamically stable (with no ZPE correction included) at pressures above 130 GPa, but is extremely close to the convex hull at all pressures and for this reason it was previously reported to have three ranges of stability < 20 GPa, 140-190 GPa and > 215 GPa.[19] In our calculations polyethylene $(CH_2)_\infty$ at some pressures is just 0.2-0.6 meV/atom above the convex hull. When taking into account zero-point energy, polyethylene turns out to have a very narrow stability region on the phase diagram (Fig 2a, b, c) at different temperatures.[5,8]

Also it was shown that two phases of bulk **graphane** with $P\text{-}3m$ (corresponds to "chair" conformation of its single unit – 6-membered carbon ring) and $Cmca$ (corresponds to "boat" conformation) space groups are thermodynamically stable (graphane I and III in ref. 15, if vdW correction is not taken into account. In this case, both these phases have no imaginary phonons frequencies and exist at 7-18 and 18-50 GPa at 0 K, respectively. In case vdW correction is included, $P\text{-}3m$ phase is unstable.

In the experimental work of Somayazulu **$4CH_4 * 2H_2$** co-crystal with space group $I4/mcm$ was reported to be stable in the range of 5.4-30 GPa.[18] In our work we also obtained a structure with the same C/H ratio and space group $I4/m$ (considering all molecules as spheres - Fig. 2b) or $I4/mcm$ (if only methane centers of mass are considered, which is reasonable due to the fact that in X-Ray diffraction experiments positions of hydrogen atoms at high pressure are not determined well). This structure is stable from 2 GPa up to ~90 GPa at 0 K and remains stable even at high temperatures (Fig. 1 and 3) up to about 150 GPa and 2000 K (at 3000 K only diamond and hydrogen are stable), when methane turns out to be not stable anymore.

We also found two new methane-hydrogen co-crystals: **$2CH_4 * 7H_2$** and **$CH_4 * 9H_2$**. $2CH_4 * 7H_2$ adopts $P\text{-}3$ space group, or $P\text{-}3m1$ if $H_2$ and $CH_4$ have spherical symmetries (Fig. 2i) and $CH_4 * 9H_2$ – $P\text{-}1$ space group, or $P6_3mc$ considering spherical symmetry of all molecules (Fig. 2j). Both structures were initially found with SCAN functional, then these were re-calculated using optB88-vdw and PBE functionals. When ZPE is included $2CH_4 * 7H_2$ is located on the convex hull. We expect experimental formation of this compound to be possible. Potentially $2CH_4 * 7H_2$ can be used as a very effective hydrogen-storage material due to its wide temperature and low-pressure stability ranges and high hydrogen content, about 30 %mass (a little less, than H4M $(CH_4 * 4H_2)$[18] – the most hydrogen-rich chemical compound known for now). $CH_4 * 9H_2$ becomes unstable, when ZPE added.

A special point is a study of **methane**. Fig. 3 shows total $p – T$

phase diagram of $CH_4$. According to our results, methane undergoes the following transformations with increasing pressure at T=0 K:

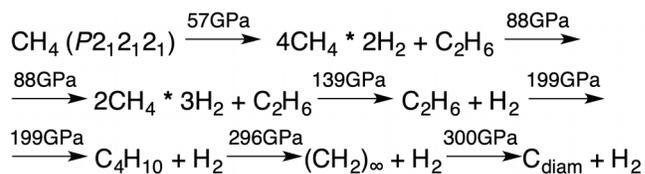

We found phase of methane with $P2_1/c$ space group (Fig. 2a) to be stable in the pressure range up to 8 GPa at finite temperatures, this result does not correspond to one of the previously published works.[37] If vdW correction isn't included this phase is stable even at 0K up to 8GPa.

In contradiction with previous works,[5,19] we show that starting from 60 GPa pure methane becomes unstable at 0 K, so that $Pnma$ phase is not on the phase diagram anymore at any temperature and pressure, since it decomposes to $4CH_4 * 2H_2$, $2CH_4 * 3H_2$ and ethane (as seen from Fig. 3). All transition pressures between different phases on $p – T$ phase diagram were re-calculated more precisely than it was done previously. Naumov and Hemley in their work obtained similar results for $CH_4$ as we did at 200 and 300 GPa, but different at 100 GPa, since $2CH_4 * 3H_2$ phase was not taken into account.[8] Appearance of $2CH_4 * 3H_2$ and $4CH_4 * 2H_2$ co-crystals on the $p – T$ phase diagram is also a new result of our work.

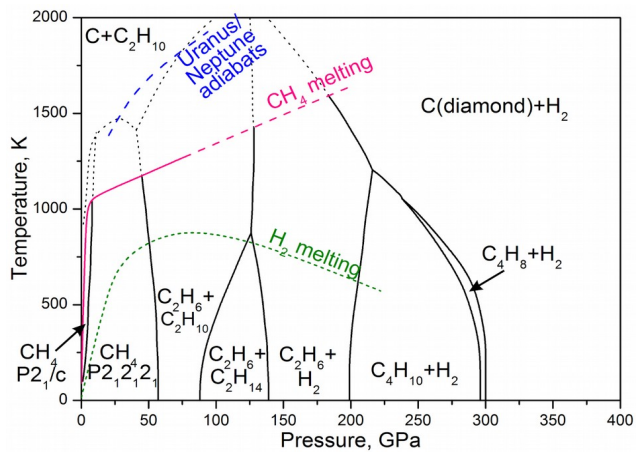

Figure 3. The phase diagram of $CH_4$ as a function of pressure and temperature. The melting line of $CH_4$ comes from Ref. 38 and $H_2$ from Ref. 39, and Uranus/Neptune adiabats are in Ref. 40, dashed lines indicate melting line approximations by Simon-Glatzel equation.[41]

In conclusion, we have shown that for C-H system only saturated hydrocarbons – alkanes – are thermodynamically stable, there are no stable unsaturated compounds. We have studied full phase diagram of the C-H system in wide ranges of temperature (up to 3000 K) and pressure (up to 400 GPa) and found that several methane-hydrogen co-crystals are thermodynamically stable. Several new structures were explored during our studies: inclusion compound $2CH_4 * H_2$, earlier obtained experimentally, discovered to have $I4/m$ space group and 2-90 GPa stability range at 0 K. $P2_1/c$ phase of methane was discovered to be stable at low pressures; bulk graphane (CH) was shown to be thermodynamically stable at 18 – 50 GPa and 0 K with *Cmca* space group; polyethylene was found on the phase diagram in a narrow range. Two new methane hydrogenates 2CH$_4$ * 7H$_2$ (*P*-3*m*1 space group) and CH$_4$ * 9H$_2$ (*Cm* space group) were discovered, that can possibly be used for hydrogen storage purpose.

# AUTHOR INFORMATION

## Corresponding Author

*E-mail: naumova.nastasiya@gmail.com

# ACKNOWLEDGMENT

Calculations were performed on our Rurik supercomputer, the Arkuda supercomputer of Skolkovo Foundation, and at the supercomputer center of the University of Nizhny Novgorod. We acknowledge Valery V. Roizen and Efim A. Mazhnik for their kind methodological advices.
We thank the Russian Science Foundation (grant 19-72-30043) and Russian Foundation for Basic Research (grant 19-02-00394) for supporting this work.